\documentclass{pasj00}

\begin{document}
\SetRunningHead{K. Sugiyama et al.}{6.7~GHz Methanol Masers with JVN}
\Received{2007/05/16}
\Accepted{2007/09/10}

\title{Mapping Observations of 6.7~GHz Methanol Masers\\
with Japanese VLBI Network}



%
\author{Koichiro \textsc{Sugiyama},\altaffilmark{1}
        Kenta \textsc{Fujisawa},\altaffilmark{2} 
        Akihiro \textsc{Doi},\altaffilmark{3,2}
        Mareki \textsc{Honma},\altaffilmark{4,5}
        Hideyuki \textsc{Kobayashi},\altaffilmark{4,6}\\
        Takeshi \textsc{Bushimata},\altaffilmark{4,7}
        Nanako \textsc{Mochizuki},\altaffilmark{3}
        and
        Yasuhiro \textsc{Murata}\altaffilmark{3,8}
        }
  \altaffiltext{1}{Graduate school of Science and Engineering,
Yamaguchi University,\\1677-1 Yoshida, Yamaguchi, Yamaguchi 753-8512}
  \altaffiltext{2}{Department of Physics, Faculty of Science,
Yamaguchi University,\\1677-1 Yoshida, Yamaguchi, Yamaguchi 753-8512}
  \altaffiltext{3}{The Institute of Space and Astronautical Science,
Japan Aerospace Exploration Agency,\\3-1-1 Yoshinodai, Sagamihara,
Kanagawa 229-8510}
  \altaffiltext{4}{VERA Project, National Astronomical Observatory
of Japan, 2-21-1 Osawa, Mitaka, Tokyo 181-8588}
  \altaffiltext{5}{Department of Astronomical Science, Graduate
University for Advanced Studies,\\2-21-1 Osawa, Mitaka, Tokyo 181-8588}
  \altaffiltext{6}{Mizusawa VERA Observatory, 2-12 Hoshigaoka,
Mizusawa, Iwate 023-0861}
  \altaffiltext{7}{Space VLBI Project, National Astronomical
Observatory of Japan, 2-21-1 Osawa, Mitaka, Tokyo 181-8588}
  \altaffiltext{8}{Department of Space and Astronautical Science,
The Graduate University for Advanced Studies,\\3-1-1 Yoshinodai,
Sagamihara, Kanagawa 229-8510}
\email{j011vc@yamaguchi-u.ac.jp}


\maketitle

\begin{abstract}

We have observed 13 methanol maser sources associated with massive
star-forming regions;
W3(OH), Mon~R2, S~255, W~33A, IRAS~18151$-$1208, G~24.78$+$0.08,
G~29.95$-$0.02, IRAS~18556$+$0136, W~48, OH~43.8$-$0.1, ON~1, Cep~A
and NGC~7538 at 6.7~GHz using the Japanese VLBI Network (JVN).
Twelve of the thirteen sources were detected at our longest baseline
of $\sim $50~M$\lambda $, and their images are presented.
Seven of them are the first VLBI images at 6.7~GHz.
This high detection rate and the small fringe spacing of
$\sim$4~milli-arcsecond suggest that most of the methanol maser
sources have compact structure.
Given this compactness as well as the known properties of long-life
and small internal-motion, 
this methanol maser line is suitable for astrometry with VLBI.

\end{abstract}


\section{Introduction}\label{section:introduction}
The class II methanol masers are well-known as tracers of early stages
of high-mass star formation (\cite{1998MNRAS.301..640W};
\cite{2001A&A...369..278M}; \cite{2006ApJ...638..241E}).
Classes I and II are defined on the basis of associated sources
(\cite{1987Natur.326...49B}; \cite{1991ASPC...16..119M});
different pumping mechanisms are proposed for them
(\cite{1992MNRAS.259..203C}, \cite{1997MNRAS.288L..39S}).
The class II masers are represented by 6.7 and 12.2~GHz lines
(\cite{1991ApJ...380L..75M}; \cite{1987Natur.326...49B}).
The spot size of class II masers is several AU
(\cite{1992ApJ...401L..39M}; \cite{1999ApJ...519..244M}),
while the class I masers are resolved out with Very Long
Baseline Interferometric (VLBI) technique \citep{1998AAS...193.7101L}.

VLBI observations for astrometry with hydroxyl, water and methanol
masers have been made using the phase referencing technique.
The accuracy of measuring with hydroxyl masers is
$\sim$1~milli-arcsecond (mas) \citep{2003A&A...407..213V},
while that achieved with water masers is up to a few tens of
micro-arcsecond ($\mu $as) \citep{2006ApJ...645..337H}.
\citet{2006Sci...311...54X} also have achieved the positional
accuracy of $\sim $10~$\mu $as with methanol maser at 12.2~GHz.
This observation has been the only one astrometric observation
with methanol masers.
Internal proper motions are often measured typically at a few mas
per year for water masers at 22~GHz
(e.g., \cite{1981ApJ...247.1039G}, \yearcite{1981ApJ...244..884G})
and this motion sometimes prevents the separation of the annual
parallax from the internal proper motion.
The internal proper motion for class II methanol masers are small
and have been measured only for W3(OH) at 12.2~GHz
\citep{2002ApJ...564..813M}.
The lifetime of 22~GHz water masers is sometimes too short for
measuring annual parallax.
For example, several spots disappear over timescales of 1 month for
Cepheus~A and IRAS~21391$+$5802 (\cite{2001ApJ...560..853T},
\yearcite{2001Natur.411..277T}; \cite{2000ApJ...538..268P}).
A monitoring of variability of 6.7~GHz methanol line for four years
showed that each spectral feature survives regardless of some
variability \citep{2004MNRAS.355..553G},
namely the lifetime of this maser is usually long enough
for measuring annual parallax.
There are 519 sites of 6.7~GHz methanol maser in a list
compiled by \citet{2005A&A...432..737P}.
Recently, new 48 sources have been detected using the 305~m Arecibo
radio telescope \citep{2007ApJ...656..255P}.
For the above reasons, i.e., compactness, small internal proper motion,
long-life and large number of known sources, class II methanol masers
may also be a useful probe for astrometry.

The VLBI Exploration of Radio Astrometry (VERA;
\cite{Kobayashi_etal.2003}) is a dedicated VLBI network for astrometry,
which mainly observes galactic water masers
for measuring their distance and motion.
Methanol masers at 6.7~GHz would also contribute to revealing
the Galactic structure as well as water masers.
The spot size of methanol maser should be small enough
for high precision astrometry.
The study of spot size, however, have been made
only a few cases so far.
The number of sources imaged with baseline of
$\geq $50~M$\lambda$ at 6.7~GHz were only four
(\cite{1992ApJ...401L..39M}; \cite{2005A&A...442L..61B};
\cite{2006A&A...448L..57P}; \cite{2007A&A...461.1027G}).
\citet{2002A&A...383..614M} discussed the size and structure of
individual masing regions in detail based on their 12.2
and 6.7~GHz observations.
They showed that the majority of the masing regions consist of
a compact maser core surrounded by extended emission (halo),
and derived the core size of 2 to 20~AU.

We have started investigations of methanol masers using the Japanese
VLBI Network (JVN) for astrometry.
The network is a newly-established one with 50--2560~km baselines
across the Japanese islands \citep{2006astro.ph.12528D}
and consists of ten antennas,
including four radio telescopes of the VERA.
We have installed 6.7~GHz receivers on three telescopes of the JVN
and have made a snap-shot imaging survey toward thirteen sources
associated with massive star-forming regions.
The aims of this observation are to investigate the detectability
of 6.7~GHz masers with our longest baseline of $\sim$50~M$\lambda$
corresponding fringe spacing of $\sim$4~mas, and to verify the imaging
capability of this network.

In this paper, we describe the detail of this observation and
data reduction in section \ref{section:observation}.
In section \ref{section:result} we present the images of the
detected sources and report the results on each individual source.
Finally, we discuss the properties of this maser for astrometry
in section \ref{section:discussion},
based on the results of this observation.


\begin{table*}
\caption{The sample of 6.7~GHz methanol masers}\label{tab:table1}
\begin{center}
\begin{tabular}{lllcrcc} 
\hline\hline
Source & \multicolumn{2}{c}{Coordinates(J2000)}                & Ref.
& \multicolumn{1}{c}{$S_{\mathrm{p}}$} &   $d$   
& \multicolumn{1}{c}{VLBI obs.} \\
       & \multicolumn{1}{c}{RA}      & \multicolumn{1}{c}{Dec} &     
&                                      &         
& \multicolumn{1}{c}{ }         \\
       & \multicolumn{1}{c}{(h m s)} 
& \multicolumn{1}{c}{($ ^{\circ} $ $ ^{\prime} $ $ ^{\prime\prime} $)}
&         & \multicolumn{1}{c}{(Jy)}  & (kpc) & \\\hline
W3(OH)              & 02 27 03.820 & \hspace{1.5mm} 61 52 25.40  & 10 
& 3294  & \hspace{0.8mm} 1.95 &    1   \\
Mon~R2              & 06 07 47.867 &             $-$06 22 56.89  &  4 
&  104  & \hspace{0.8mm} 0.83 & 6, 7   \\
S~255               & 06 12 54.024 & \hspace{1.5mm} 17 59 23.01  &  6 
&   79  & 2.5  & 6, 7   \\
W~33A               & 18 14 39.52  &             $-$17 51 59.7   & 13 
&  297  & 4.0  &  \ldots \\
IRAS~18151$-$1208   & 18 17 58.07  &             $-$12 07 27.2   & 13 
&  119  & 3.0  &  8   \\
G~24.78$+$0.08 & 18 36 12.57 & $-$07 12 11.4   & \footnotemark[$*$] 
&   84  & 7.7  & \ldots \\
G~29.95$-$0.02      & 18 46 03.741 &             $-$02 39 21.43  &  6 
&  182  & 9.0  & \ldots \\
IRAS~18556$+$0136   & 18 58 13.1   & \hspace{1.5mm} 01 40 35     & 12 
&  191  & 2.0  & \ldots \\
W~48                & 19 01 45.5   & \hspace{1.5mm} 01 13 28     &  3 
&  733  & 3.4  &   6   \\
OH~43.8$-$0.1       & 19 11 53.987 & \hspace{1.5mm} 09 35 50.308 &  2 
&   51  & 2.8  & \ldots \\
ON~1                & 20 10 09.1   & \hspace{1.5mm} 31 31 34     & 12 
&  107  & 1.8  & \ldots \\
Cep~A               & 22 56 17.903 & \hspace{1.5mm} 62 01 49.65  & 13 
&  371  & \hspace{0.8mm} 0.73 & \ldots \\
NGC~7538            & 23 13 45.364 & \hspace{1.5mm} 61 28 10.55  &  6 
&  256  & 2.8  &  5, 6, 7, 9, 11 \\\hline
\multicolumn{7}{l}{\hbox to 0pt{\parbox{136mm}{\footnotesize
Col~(1)~source name; Col~(2)--(3)~coordinates in J2000;
Col~(4)~reference of coordinates;
Col~(5)~peak flux densities from the single-dish observations
by Yamaguchi 32~m; Col~(6)~source distance; Col~(7)~Published
VLBI observations at 6.7~GHz.

References --- (1)~\cite{1992ApJ...401L..39M};
(2)~\cite{1994ApJS...91..659K}; (3)~\cite{1995MNRAS.272...96C};
(4)~\cite{1998MNRAS.301..640W}; (5)~\cite{1998A&A...336L...5M};
(6)~\cite{2000A&A...362.1093M}; (7)~\cite{2001A&A...369..278M};
(8)~\cite{2002evn..conf..213V}; (9)~\cite{2004ApJ...603L.113P};
(10)~\cite{2005MNRAS.360.1162E}; (11)~\cite{2006A&A...448L..57P};
(12)~\cite{1994yCat.2125....0J}; (13)~Fringe rate mapping in our
observations (accuracy of position is from 100 to 300~mas).

\footnotemark[$*$] The coordinate of this source is not used for
this data reduction. This coordinate is obtained from the
observations made in the following year.
It is coincident with the coordinate in the catalog
listed by \citet{2005A&A...432..737P}.}\hss}}
\end{tabular}
\end{center}
\end{table*}


\section{Observations and Data Reduction}\label{section:observation}
\subsection{VLBI Observation}\label{section:VLBI}
The $ 5_1\rightarrow6_0A^+ $ methanol transition at 6668.518~MHz was
observed on 2005 September 26 from 5:00 to 21:00 UT using three
telescopes (Yamaguchi 32~m, VERA-Mizusawa 20~m, VERA-Ishigaki 20~m)
of the JVN.
The maximum fringe spacing was 9.1~mas (Yamaguchi--Mizusawa,
22~M$\lambda $) and the minimum was 4.1~mas
(Mizusawa--Ishigaki, 50~M$\lambda $).
Right-circular polarization was received at Yamaguchi
with system noise temperature of 220~K, 
while linear polarization was received at Mizusawa and Ishigaki
stations with system noise temperatures of 120~K.
The data were recorded on magnetic tapes using the VSOP-terminal
system at a data rate of 128~Mbps with 2-bit quantization
and 2 channels, and correlated at the Mitaka FX correlator
\citep{Shibata_etal.1998}.
From the recorded 32~MHz bandwidth, 8~MHz (6664~MHz to 6672~MHz)
was divided into 1024 channels and used for data analysis,
yielding a velocity resolution of 0.35~km~s$^{-1}$.

Sources that are bright and have widespread velocity range were
selected as targets of this observation.
The following thirteen sources were observed; W3(OH), Mon~R2, S~255,
W~33A, IRAS~18151$-$1208, G~24.78$+$0.08, G~29.95$-$0.02,
IRAS~18556$+$0136, W~48, OH~43.8$-$0.1, ON~1, Cep~A and NGC~7538.
The details of each source are shown in table~\ref{tab:table1}.
Scans of 15~minutes duration were made for 2-4 times on each source
at different hour angles in order to improve uv-coverages.
The number of scans for each source is shown in table~\ref{tab:table2}.
Strong continuum sources, NRAO~530, 3C~454.3, and 3C~84, were observed
every two hours for bandpass and delay calibration.

The data were reduced using the Astronomical Image Processing System
(AIPS; \cite{Greisen2003}).
Correlator digitization errors were corrected using the task ACCOR.
Clock offsets and clock rate offsets were corrected
using strong continuum calibrators in the task FRING.
Bandpass calibration was performed using strong continuum calibrators
in the task BPASS.
Doppler corrections were made by running the tasks SETJY and CVEL.
Amplitude calibration parameters were derived from the total-power
spectra of maser lines using template-method in the task ACFIT.
Fringe-fitting was conducted using one spectral channel including
a strong maser feature in the task FRING.
The solutions were applied to all the other channels.

We have searched maser spots over an area of
\timeform{4".0} $\times $ \timeform{4".0} with the Difmap software
\citep{Shepherd1997}.
Structure models were made by model-fitting with point sources and
self-calibration algorithms iteratively.
The phase solutions of self-calibration were applied
to all the other channels.

In addition to the analyses described above, special amplitude
calibrations were necessary because different polarizations
(circular/linear) were correlated.
Since a visibility amplitude reduces by $1/\sqrt{2} $ in correlation
between linear and single circular polarization,
the factor was corrected.
For the case of linear and linear polarization,
the amplitude varies with time depending
on position angle between antennas.
Amplitude correction factors were calculated for each observational
scan and applied to the visibilities.
We made a baseline-based correction to each observational scan.
The correction procedure was confirmed in some observations
at different date by applying the task BLCAL
for bright continuum sources.
This process indicated that an accuracy of this calibration
was $\sim $10~{\%}.

\subsection{Spectroscopic Single-Dish Observation}
\label{section:single}
We also have made a series of single-dish observation
of 6.7~GHz methanol maser with the Yamaguchi~32~m telescope.
The single-dish observations were made about one month
before the VLBI observation (August 2005).
In this paper, we use the spectra as the total-power
(not cross-correlated) spectra in comparison
with cross-correlated spectra.
The accuracy of correlated flux density depends on the accuracy
of the single-dish observation.

The received signal at dual circular polarizations with a bandwidth
of 4~MHz each was divided into 4096 channels,
yielding a velocity resolution of 0.044~km~s$^{-1}$.
The flux density calibration was made using an aparture efficiency
of 70~{\%} and a system noise temperature (220~K) measured
on the first day of the single-dish observation.
The accuracy of the calibration was 10~{\%}.
The rms of noise was typically 1.0~Jy, for data combined
from dual polarizations with a 14~minutes integration.
We used source positions listed in the IRAS Point Source Catalog
(PSC; \cite{1994yCat.2125....0J}).
Error in the LSR velocity was potentially to be $\pm 0.3$~km~s$^{-1}$.


\section{Results}\label{section:result}
\begin{figure*}[htb]
  \begin{center}
    \includegraphics[width=160mm,clip]{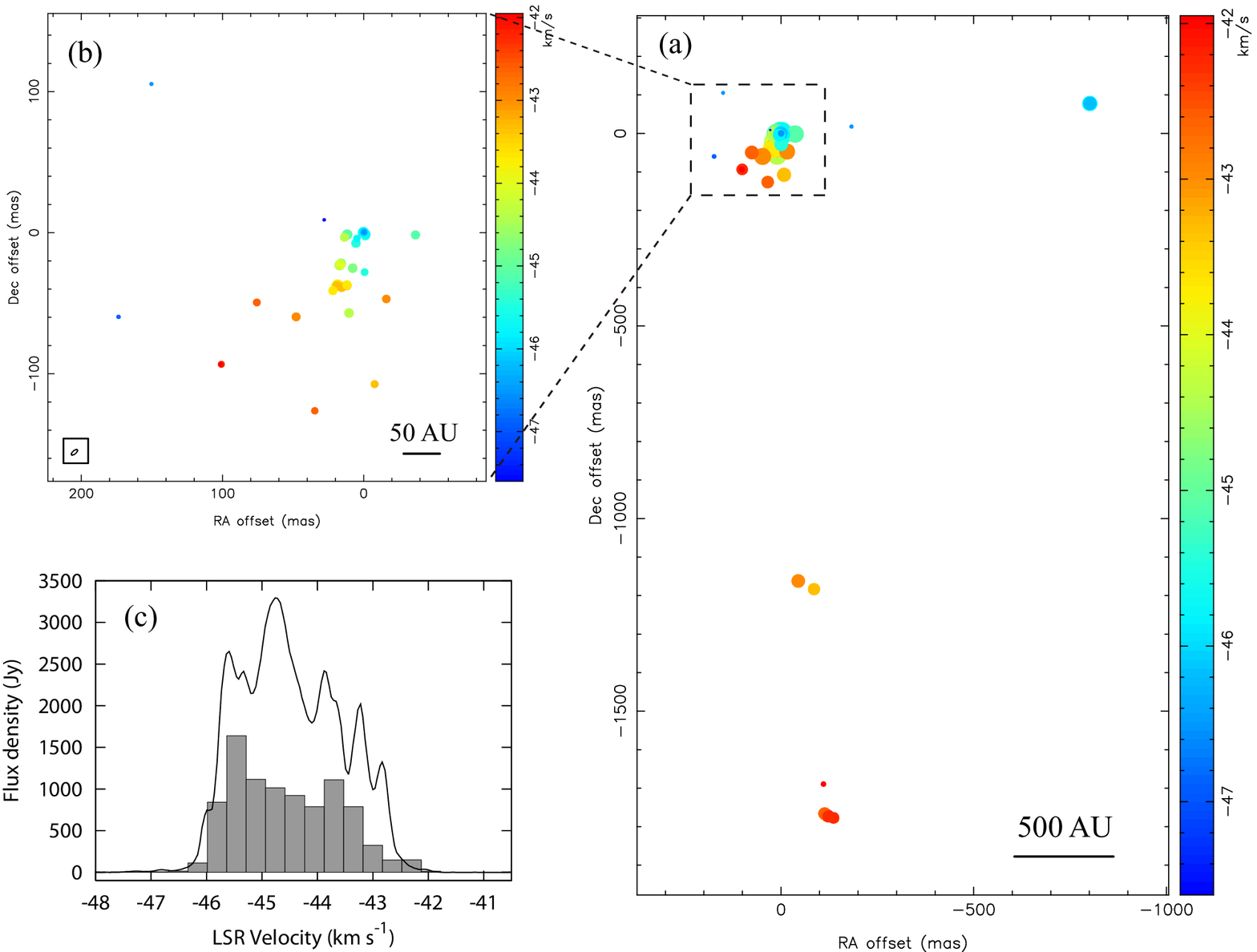}
  \end{center}
 \caption{W3(OH):
         (a) wide field view of
\timeform{0".9}~$\times$~\timeform{1".8} channel-velocity map.
         (b) close-up of the main cluster.
		 (c) CLEANed component spectrum (filled block)
and total spectrum (solid curve).
		  }
 \label{fig:fig1}
\end{figure*}

\begin{figure*}[htb]
  \begin{center}
    \includegraphics[width=160mm,clip]{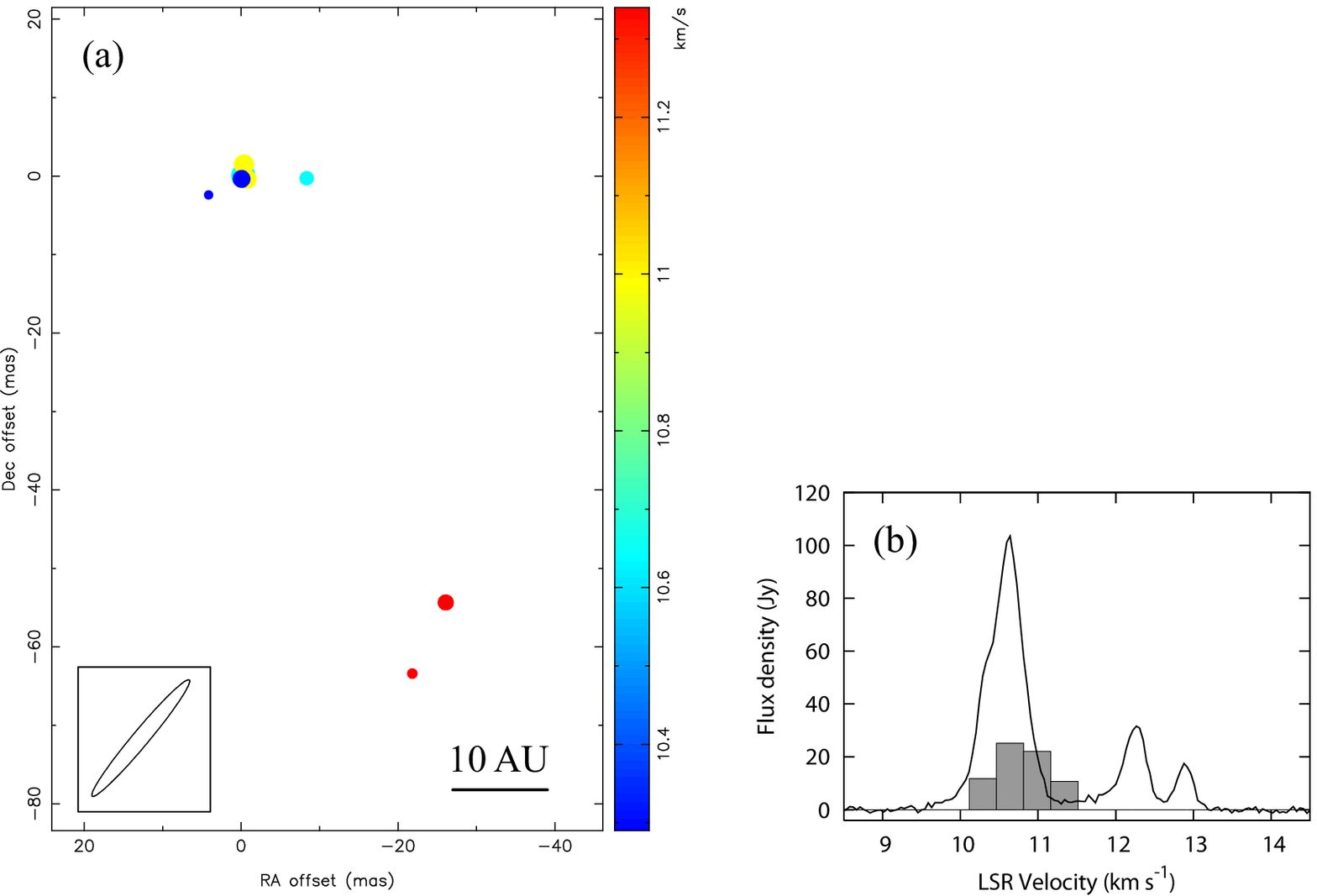}
  \end{center}
 \caption{Mon~R2:
          (a) channel-velocity map.
          (b) CLEANed component spectrum (filled block)
and total spectrum (solid curve).}
 \label{fig:fig2}
\end{figure*}

\begin{figure*}[htb]
  \begin{center}
    \includegraphics[width=80mm,clip]{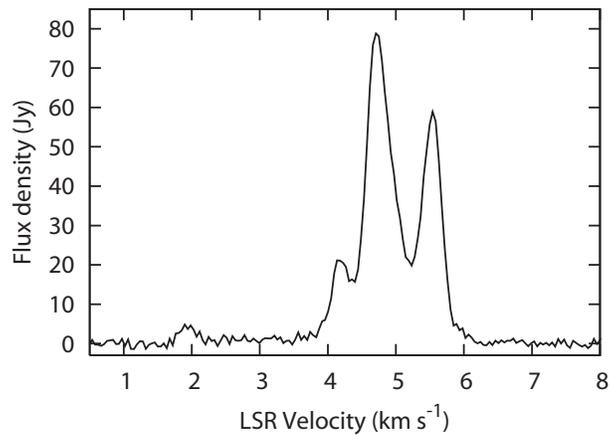}
  \end{center}
 \caption{The total spectrum of S~255 observed by Yamaguchi 32~m.
          This source was not detected by our VLBI observation.}
 \label{fig:fig3}
\end{figure*}

\begin{figure*}[htb]
  \begin{center}
    \includegraphics[width=160mm,clip]{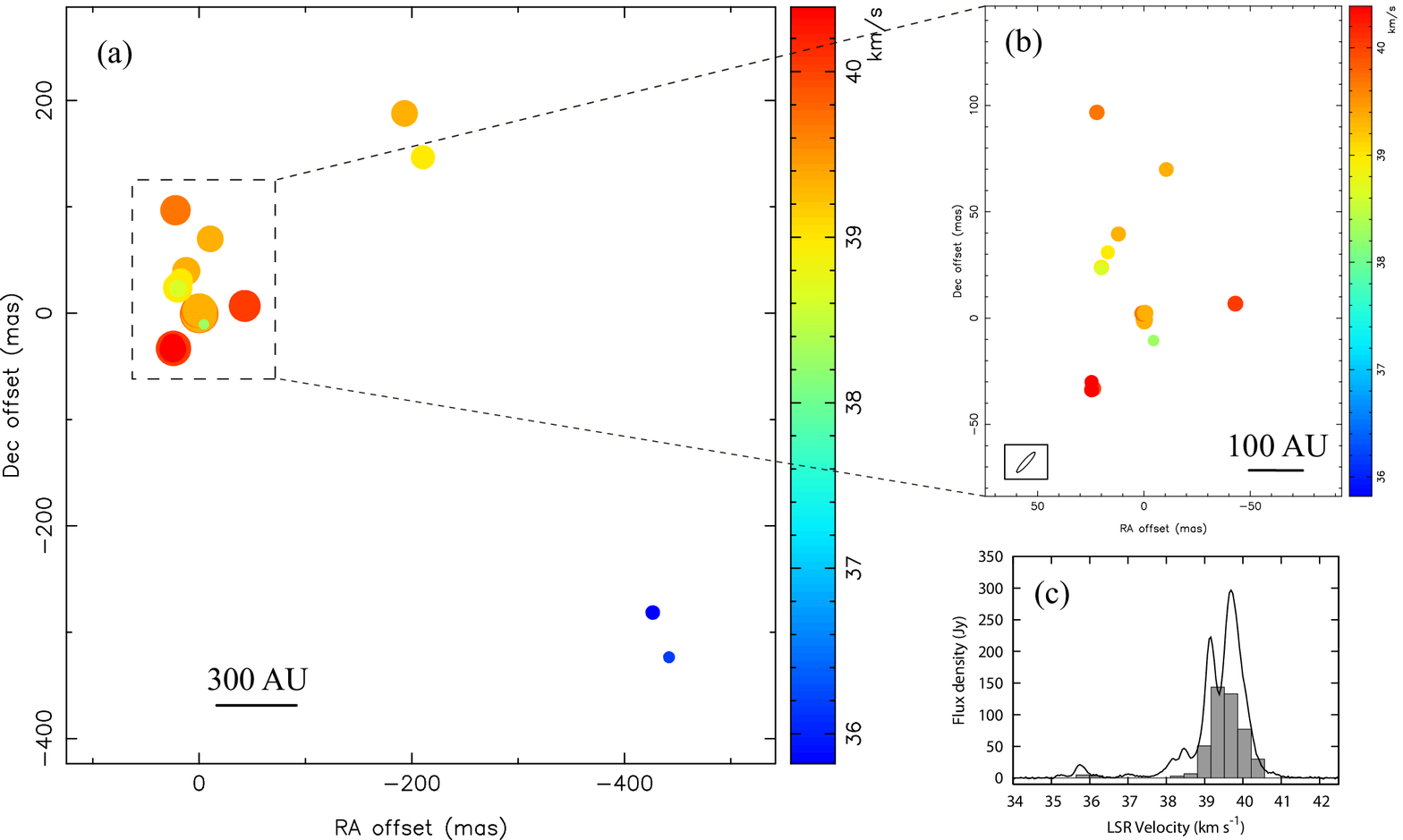}
  \end{center}
 \caption{W~33A:
          (a) wide field view of
\timeform{0".5}~$\times$~\timeform{0".5} channel-velocity map.
          (b) close-up of the main cluster.
          (c) CLEANed component spectrum (filled block)
and total spectrum (solid curve).}
 \label{fig:fig4}
\end{figure*}

\begin{figure*}[htb]
  \begin{center}
    \includegraphics[width=160mm,clip]{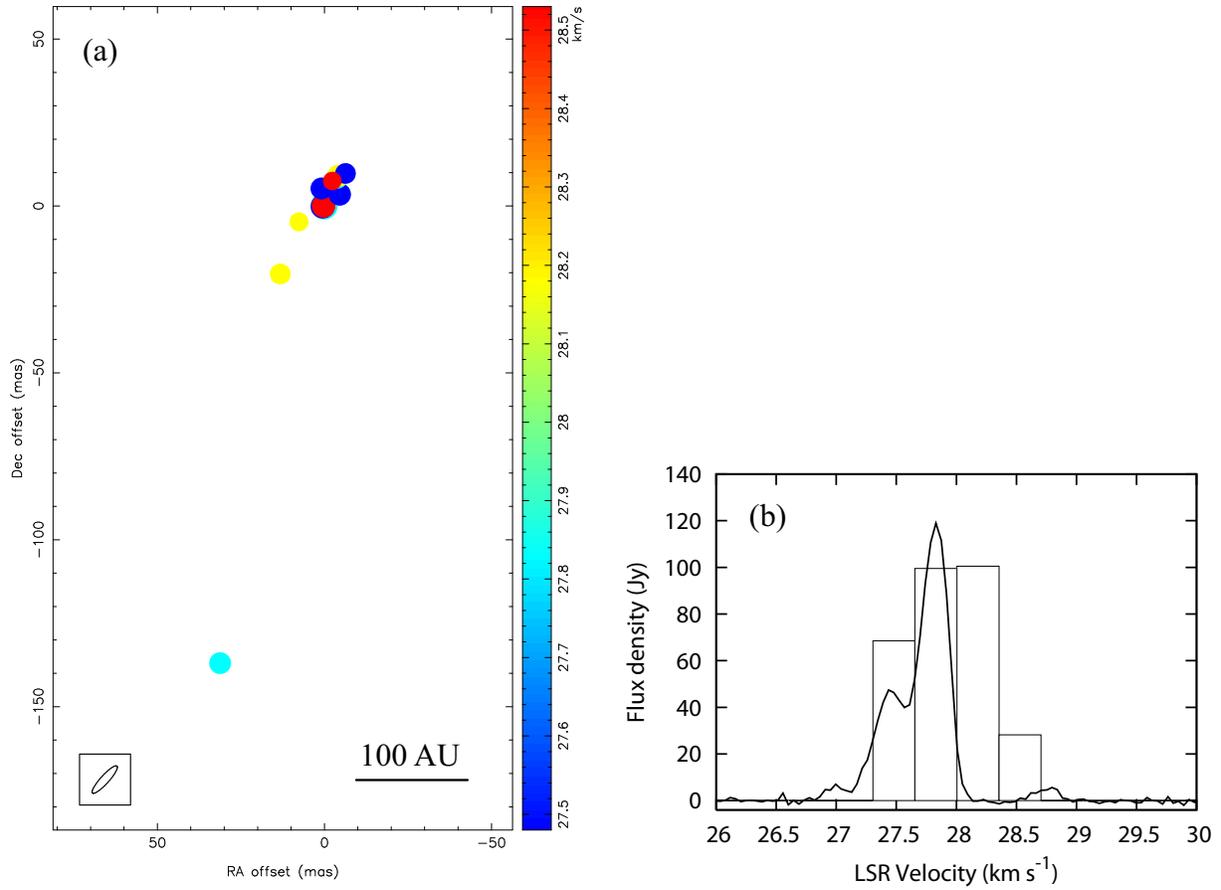}
  \end{center}
 \caption{IRAS~18151$-$1208:
          (a) channel-velocity map.
          (b) CLEANed component spectrum (blank block)
and total spectrum (solid curve).}
 \label{fig:fig5}
\end{figure*}

\begin{figure*}[htb]
  \begin{center}
    \includegraphics[width=160mm,clip]{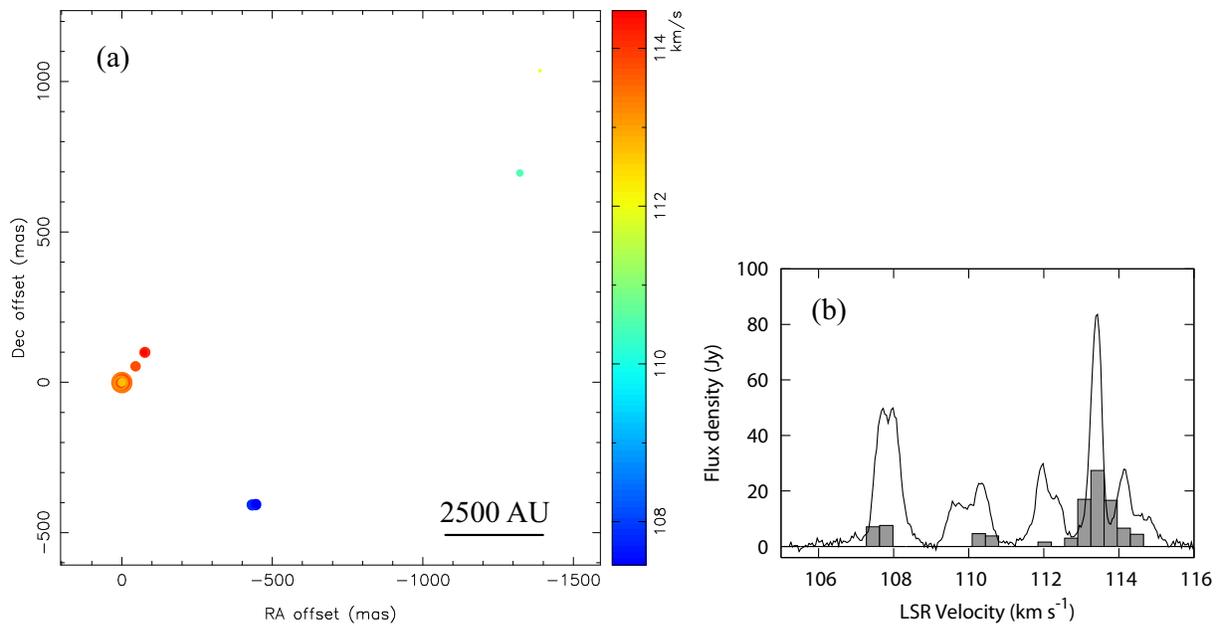}
  \end{center}
 \caption{G~24.78$+$0.08:
          (a) wide field view of
\timeform{1".4}~$\times$~\timeform{1".4} channel-velocity map.
          (b) CLEANed component spectrum (filled block)
and total spectrum (solid curve).}
 \label{fig:fig6}
\end{figure*}

\begin{figure*}[htb]
  \begin{center}
    \includegraphics[width=160mm,clip]{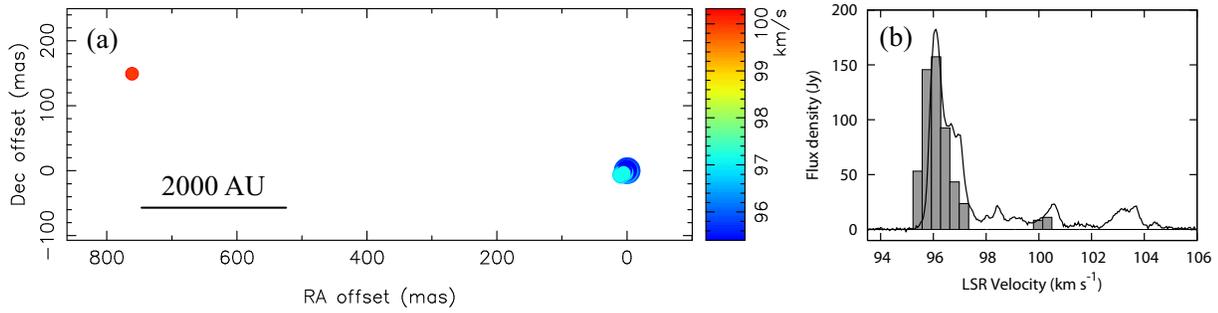}
  \end{center}
 \caption{G~29.95$-$0.02:
          (a) wide field view of
\timeform{0".76}~$\times$~\timeform{0".15} channel-velocity map.
          (b) CLEANed component spectrum (filled block)
and total spectrum (solid curve).}
 \label{fig:fig7}
\end{figure*}

\begin{figure*}[htb]
  \begin{center}
    \includegraphics[width=160mm,clip]{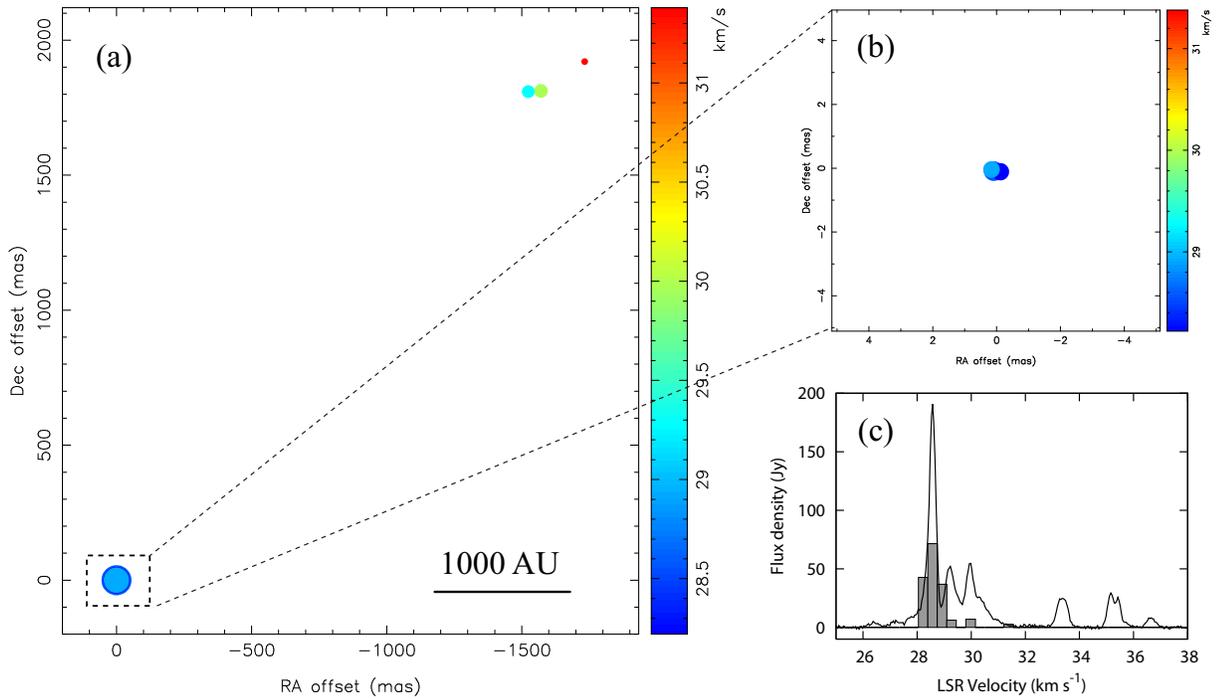}
  \end{center}
 \caption{IRAS~18556$+$0136:
          (a) wide field view of
\timeform{1".7}~$\times$~\timeform{1".9} channel-velocity map.
          (b) CLEANed component spectrum (filled block)
and total spectrum (solid curve).}
 \label{fig:fig8}
\end{figure*}

\begin{figure*}[htb]
  \begin{center}
    \includegraphics[width=160mm,clip]{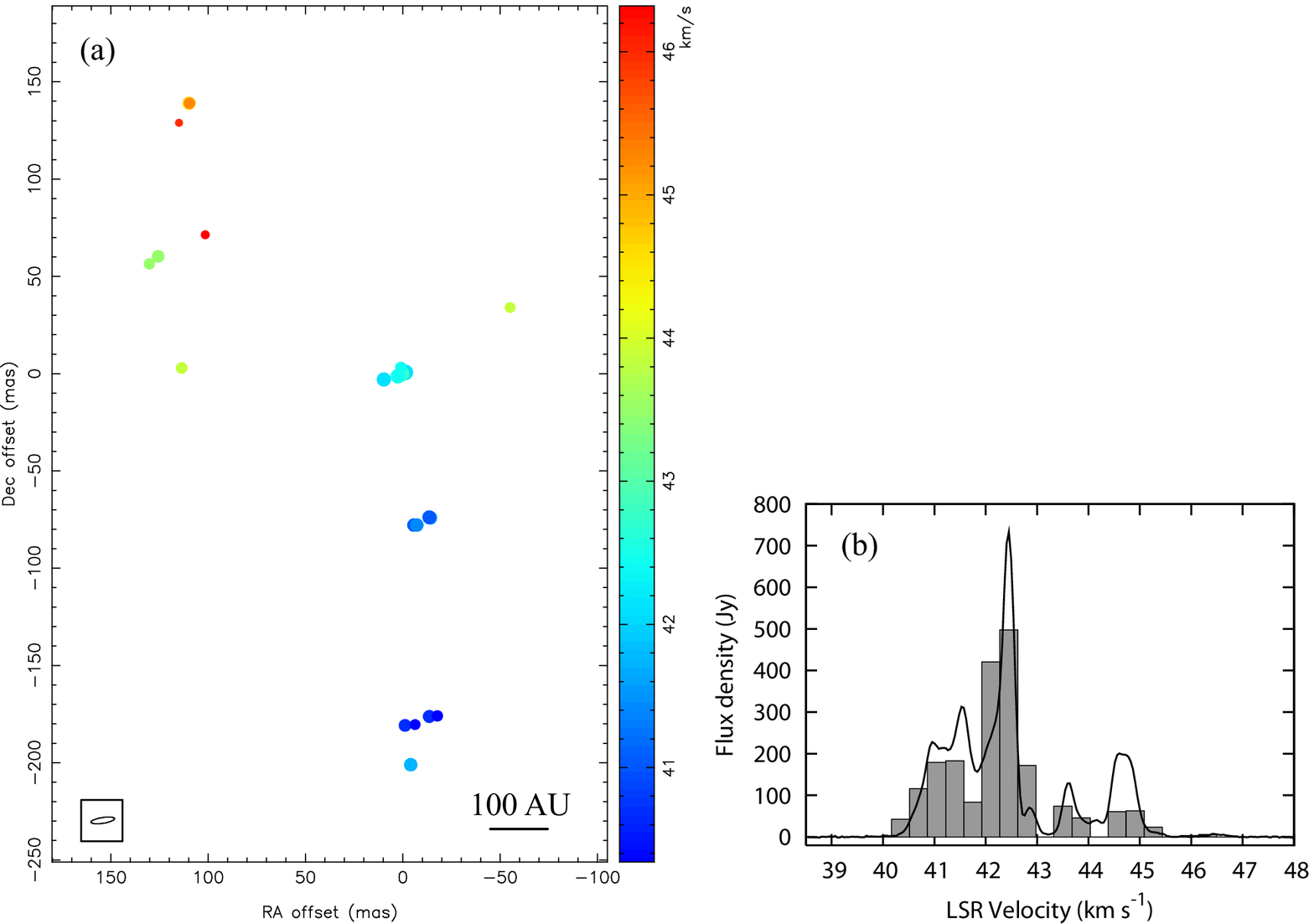}
  \end{center}
 \caption{W~48:
          (a) channel-velocity map.
          (b) CLEANed component spectrum (filled block)
and total spectrum (solid curve).}
 \label{fig:fig9}
\end{figure*}

\begin{figure*}[htb]
  \begin{center}
    \includegraphics[width=160mm,clip]{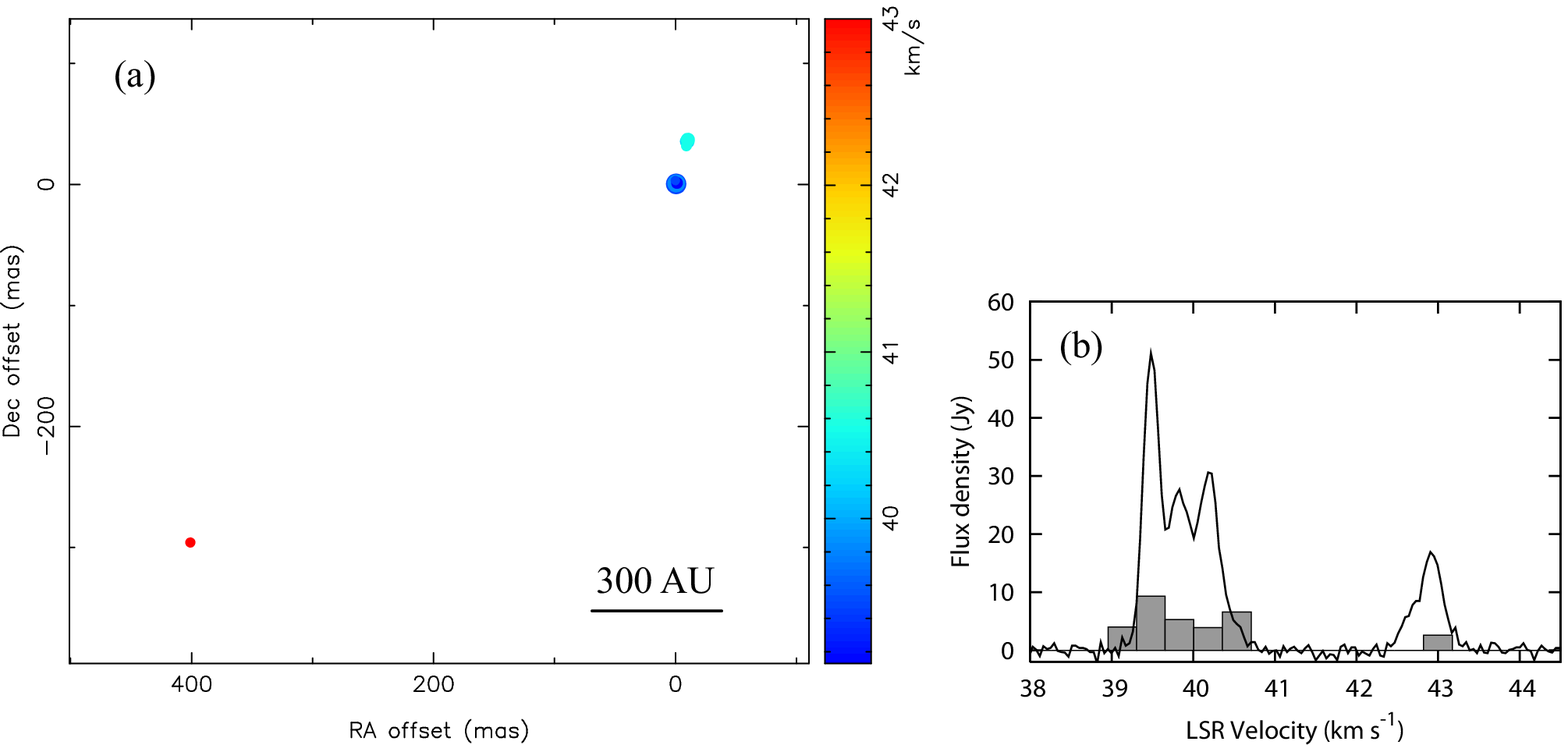}
  \end{center}
 \caption{OH~43.8$-$0.1:
          (a) wide field view of
\timeform{0".4}~$\times$~\timeform{0".3} channel-velocity map.
          (b) CLEANed component spectrum (filled block)
and total spectrum (solid curve).}
 \label{fig:fig10}
\end{figure*}

\begin{figure*}[htb]
  \begin{center}
    \includegraphics[width=160mm,clip]{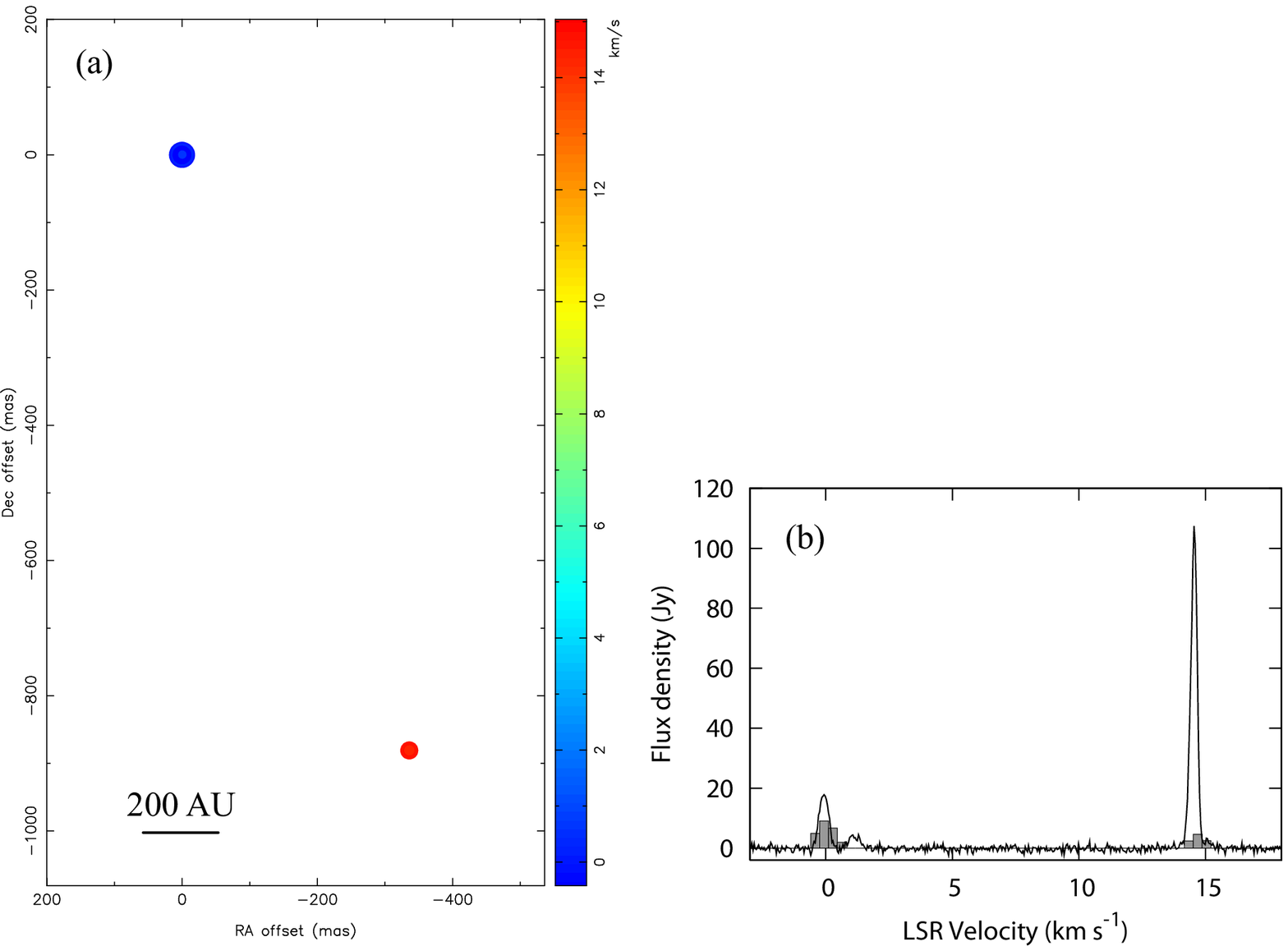}
  \end{center}
 \caption{ON~1:
          (a) wide field view of
\timeform{0".34}~$\times$~\timeform{0".88} channel-velocity map.
          (b) CLEANed component spectrum (filled block)
and total spectrum (solid curve).}
 \label{fig:fig11}
\end{figure*}

\begin{figure*}[htb]
  \begin{center}
    \includegraphics[width=160mm,clip]{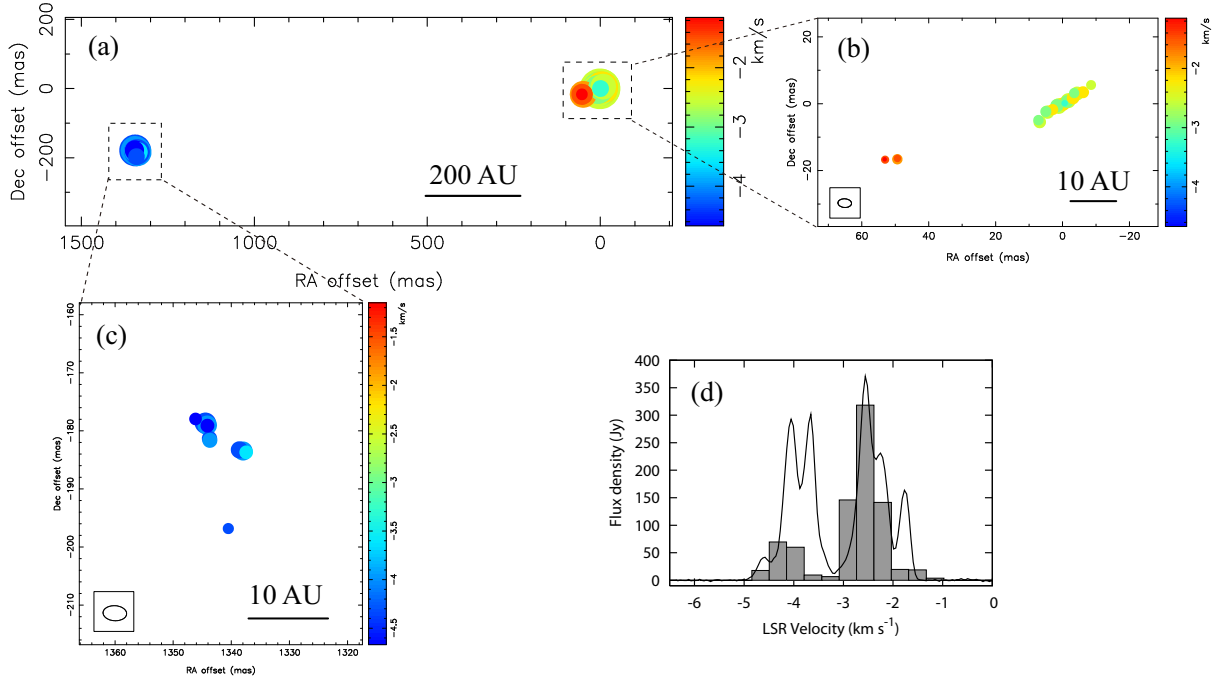}
  \end{center}
 \caption{Cep~A:
          (a) wide field view of
\timeform{1".3}~$\times$~\timeform{0".18} channel-velocity map.
          (b) close-up of the redshifted cluster.
          (c) close-up of the blueshifted cluster.
		  (d) CLEANed component spectrum (filled block)
and total spectrum (solid curve).}
 \label{fig:fig12}
\end{figure*}

\begin{figure*}[htb]
  \begin{center}
    \includegraphics[width=160mm,clip]{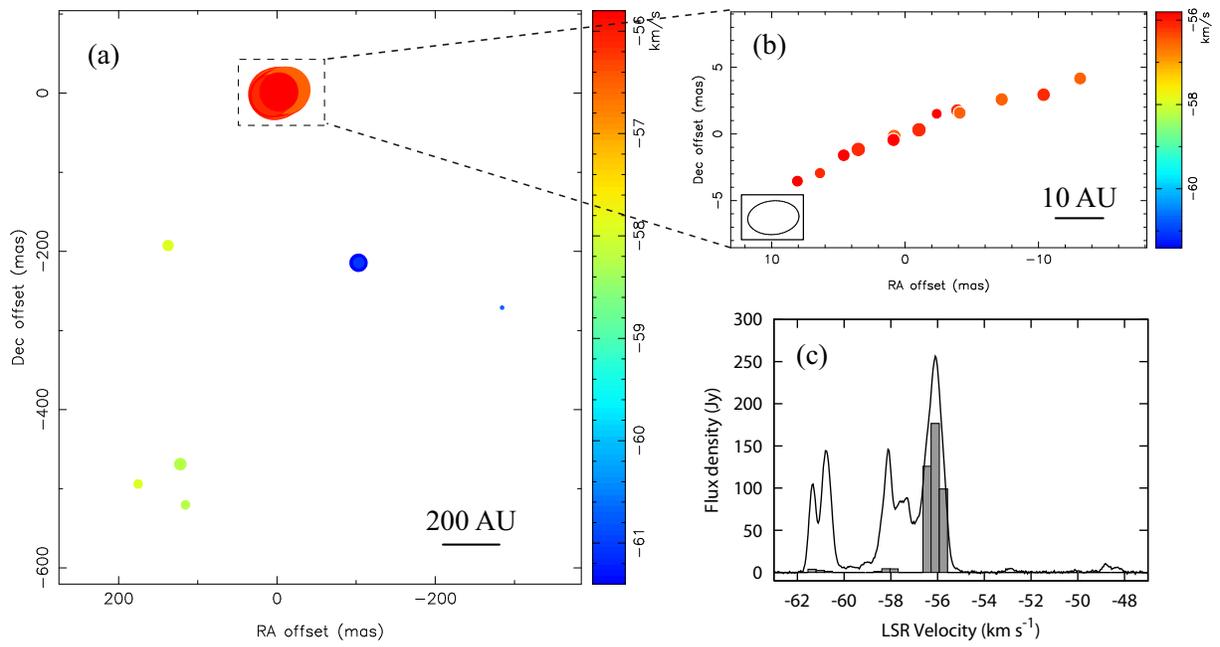}
  \end{center}
 \caption{NGC~7538:
          (a) wide field view of
\timeform{0".5}~$\times$~\timeform{0".5} channel-velocity map.
          (b) close-up of the redshifted cluster.
          (c) CLEANed component spectrum (filled block)
and total spectrum (solid curve).}
 \label{fig:fig13}
\end{figure*}

All the targets except for S~255 were detected.
The channel-velocity maps of the detected sources are presented
in figures~\ref{fig:fig1}--\ref{fig:fig13}.
Only total-power spectrum is shown in figure~\ref{fig:fig3} for S~255.
The maps indicates positions of maser spots relative to that
of the reference spot; the size and color of a spot represent
its flux density in logarithmic scale and radial velocity,
respectively.
Correlated spectra (total CLEANed components) are shown in addition
to total-power spectra for each source.
The rms of image noise in a line-free channel
ranges from 200 to 670~mJy~beam$^{-1}$.
The minimum detectable sensitivities of 7~$\sigma $ were in the range
from 1.4 to 4.7~Jy~beam$^{-1}$.
The maximum dynamic range was 196 for W3(OH).
The projected baseline ranged from 11 to 50~M$\lambda $ for all sources
except for W~33A, IRAS~18151$-$1208 and G~24.78$+$0.08
(7 to 50~M$\lambda $).
Flux ratios of correlated to total spectra at the peak channel
of total spectra, and that of the integrated spectra are shown
on column~7, 8 in table~\ref{tab:table2}, respectively.
Ratio of correlated flux densities at the longest baseline of
50~M$\lambda $ to that at zero-baseline (total-power)
was 20~{\%} on average.
We describe the results on each individual source below
in Right Ascension order.


\begin{table*}
\caption{Observational Results}\label{tab:table2}
\begin{center}	
\begin{tabular}{lcrrcrcc}\hline\hline
Source & $N_{\mathrm{scan}}$ & 
$\theta_{\mathrm{maj}} \times \theta_{\mathrm{min}}$  & $P.A.$ 
& $\sigma $                   & \multicolumn{1}{c}{$v_{\mathrm{ref}}$}
& $S_{\mathrm{VLBI}}^{\mathrm{p}}/S^{\mathrm{p}}$ 
& $S_{\mathrm{VLBI}}/S$   \\
       &                     & 
(mas $\times$ mas)                                    & (deg)  
& $ (\textrm{Jy~beam}^{-1}) $ & $ (\textrm{km~s}^{-1}) $ 
& (\%) & (\%)  \\\hline
W3(OH)              & 3 & $  5.5 \times 2.3 $ & 131 &   0.24  
& $-$45.46 & 31     &     \hspace{1.7mm}45   \\
Mon~R2              & 2 & $ 19.3 \times 2.1 $ & 140 &   0.30  
&    10.64 & 24     &     \hspace{1.7mm}33   \\
S~255               & 2 & $ 19.4 \times 2.1 $ & 144 & \ldots  
&   \ldots & \ldots &     \hspace{1.7mm}\ldots  \\
W~33A               & 3 & $ 12.6 \times 2.5 $ & 138 &   0.37  
&    39.69 & 45     &     \hspace{1.7mm}51   \\
IRAS~18151$-$1208   & 3 & $ 10.9 \times 2.5 $ & 138 &   0.63  
&    27.83 & 84     &     208   \\
G~24.78$+$0.08      & 3 & $  9.5 \times 2.4 $ & 135 &   0.28  
&   113.43 & 33     &     \hspace{1.7mm}27   \\
G~29.95$-$0.02      & 2 & $ 19.5 \times 2.2 $ & 139 &   0.46  
&    96.10 & 86     &     \hspace{1.7mm}78   \\
IRAS~18556$+$0136   & 3 & $  9.4 \times 2.5 $ & 135 &   0.32  
&    28.57 & 37     &     \hspace{1.7mm}36   \\
W~48                & 3 & $ 12.2 \times 3.0 $ &  99 &   0.67  
&    42.45 & 68     &     \hspace{1.7mm}91   \\
OH~43.8$-$0.1       & 2 & $ 16.2 \times 2.2 $ & 139 &   0.36  
&    39.48 & 18     &     \hspace{1.7mm}30   \\
ON~1                & 4 & $  4.3 \times 2.5 $ & 112 &   0.23  
&  $-$0.07 & \hspace{1.6mm}4     &     \hspace{1.7mm}27   \\
Cep~A               & 4 & $  4.1 \times 2.6 $ &  87 &   0.62  
&  $-$2.56 & 86     &     \hspace{1.7mm}65   \\
NGC~7538            & 4 & $  3.9 \times 2.5 $ &  95 &   0.20  
& $-$56.10 & 69     &     \hspace{1.7mm}30   \\\hline
\multicolumn{8}{l}{\hbox to 0pt{\parbox{155mm}{\footnotesize
Col~(1)~source name; Col~(2)~number of scans;
Col~(3)~FWMHs of major and minor axes of synthesized beam;
Col~(4)~position angle of major axis of beam;
Col~(5)~rms of image noise in line-free channels;
Col~(6)~reference velocity channel; Col~(7)~flux ratio of correlated
to total spectra at the peak channel of total spectra;
Col~(8)~flux ratio of correlated to total spectra of the
integrated spectra.}\hss}}
\end{tabular}
\end{center}
\end{table*}


\subsection{W3(OH)}\label{section:W3(OH)}
W3(OH) (figure~\ref{fig:fig1}) is a well-studied star-forming region
at a distance of 1.95~$ \pm $~0.04~kpc \citep{2006Sci...311...54X}
containing a hot molecular core (HMC) and an UC H\emissiontype{II}
region \citep{1984ApJ...287L..81T}.
The central object is thought to be an O9$ - $O7 young star with an
estimated mass of $ \simeq $ $ 30~\MO $ \citep{1981ApJ...245..857D}.
W3(OH) has been imaged with the Multi-Element Radio-Linked
Interferometer Network (MERLIN, \cite{2005MNRAS.360.1162E};
\cite{2006MNRAS.tmpL..65V}; \cite{2006MNRAS.371.1550H})
and VLBI array at 6.7~GHz \citep{1992ApJ...401L..39M},
and with the VLBA at 12.2~GHz (\cite{1988ApJ...333L..83M};
\cite{1999ApJ...519..244M}, \yearcite{2002ApJ...564..813M},
\yearcite{2003ApJ...583..776M}; \cite{2006Sci...311...54X}).

In our observation, 48 maser spots were detected.
The velocity of reference feature is $-$45.46~km~s$^{-1}$
($-$45.37~km~s$^{-1}$ appeared in \cite{2005MNRAS.360.1162E}).
We detected the maser clusters 1, 5, 6 and 7 defined by
\citet{1992ApJ...401L..39M},
and found a new spot of $-$46.51~km~s$^{-1}$ at 180~mas west
from the reference feature.
There are 38 spots within a 200~mas area of the cluster 6.
The other three clusters locate \timeform{0".8} west, \timeform{1".2}
south, and \timeform{1".7} south form the main cluster, respectively.

\subsection{Mon~R2}\label{section:MonR2}
The Monoceros R2 (Mon~R2, figure~\ref{fig:fig2}) molecular cloud has a
cluster of seven bright infrared sources \citep{1976ApJ...208..390B}.
The cluster is one of the closest massive star forming regions
to the solar system at a distance of 830~pc
(\cite{1968AJ.....73..233R}; \cite{1976AJ.....81..840H}).
The 6.7~GHz methanol maser of Mon~R2 has been observed with
the Australia Telescope Compact Array
(ATCA, \cite{1998MNRAS.301..640W}) and with European VLBI Network
(EVN, \cite{2000A&A...362.1093M}, \yearcite{2001A&A...369..278M}).
Although there are several peaks in the total-power spectrum,
our VLBI observation detected only one spectral feature
around 10.64~km~s$^{-1}$ as eight maser spots.
This eight spots ($V_\mathrm{lsr}=$ 10.29 to 11.34) correspond to
'C' defined by \citet{1998MNRAS.301..640W}, and also are identified
with that five of fourteen spots detected by
\citet{2000A&A...362.1093M}.

Since this source shows an on-going flux variation,
the single-dish spectrum was largely different from
that of previous observations.
It seems that some spectral features disappeared and some others
appeared during 1992 to 2005
(\cite{1995MNRAS.272...96C}; \cite{2000A&AS..143..269S}).

\subsection{S~255}\label{section:S255}
S~255 locating at a distance of 2.5~kpc \citep{1988A&A...191...44M}
includes one UC H\emissiontype{II} region G~192.58-0.04
\citep{1994ApJS...91..659K}.
S~255 has been observed with EVN at 6.7~GHz
(\cite{2000A&A...362.1093M}, \yearcite{2001A&A...369..278M}),
and seven spots were detected with the longest projected baselines
of $\sim $30~M$\lambda $.
We observed this source and correlated with the coordinates that are
used by \citet{2000A&A...362.1093M}, (\yearcite{2001A&A...369..278M}),
but no spot was detected. 
This source might be resolved out with our baselines.
The single-dish spectrum observed by Yamaguchi 32~m is shown
in figure~\ref{fig:fig3}.

\subsection{W~33A}\label{section:W33A}
W~33A (figure~\ref{fig:fig4}) is a highly luminous object
($ L=1{\times}10^5~\LO $, \cite{1984ApJ...283..573S}) and coincide
a deeply embedded massive young stars \citep{2000ApJ...537..283V}.
The kinematic distance based on CS and $ \textrm{C}^{34}\textrm{S} $
observations is 4~kpc \citep{2000ApJ...537..283V}.
A map consisting eleven maser spots has been obtained with
the ATCA \citep{1998MNRAS.301..640W}.

Our observation, the first VLBI for this source, detected 19 maser
spots which correspond to 'F', 'G', and 'K' defined
by the ATCA observation.
The total-power spectrum having several peaks, but weak spectral
peaks were not detected in the VLBI map.
This source consists of two clusters,
and each clusters are separated larger than 1000~AU.

\subsection{IRAS~18151$-$1208}\label{section:18151}
IRAS~18151$-$1208 (figure~\ref{fig:fig5}) is embedded in a high-density
cloud \citep{1996A&AS..115...81B} and thought to be
in pre-UC H\emissiontype{II} phase \citep{2004A&A...425..981D}.
The kinematic distance based on CS line observation
is 3.0~kpc \citep{1993A&A...275...67B}.
Thirteen maser spots were detected by our observation. 
This source has been observed with the ATCA \citep{2002A&A...390..289B}
and with EVN \citep{2002evn..conf..213V}.
A spot of 27.83~km~s$^{-1}$ at 137~mas south from the main cluster
was newly detected one.

\subsection{G~24.78$+$0.08}\label{section:G24}
G~24.78$+$0.08 (figure~\ref{fig:fig6}) is a cluster of massive
protostars at a distance of 7.7~kpc \citep{1989A&A...213..339F}.
A pair of cores associated with a compact bipolar outflow have been
detected \citep{2002A&A...390L...1F}.
\citet{2004ApJ...601L.187B} have detected rotating disks associated
with high-mass YSOs by observing 1.4~mm continuum and CH$_{3}$CN
($J=$ 12--11) line emission.
Our observation provides the first VLBI image of
methanol masers at 6.7~GHz for this source.
This source consists of three clusters separated larger than 3000~AU.

\subsection{G~29.95$-$0.02}\label{section:G29}
G~29.95$-$0.02 (figure~\ref{fig:fig7}) is at a distance of 9~kpc
\citep{1989ApJ...340..265W} and coincides an NH$_{3} $ hot core,
but is offset by a few arcsec from the continuum peak of
a UC H\emissiontype{II} region \citep{1998A&A...331..709C}. 
The methanol masers of G~29.95$-$0.02 has been observed with
the ATCA at 6.7~GHz \citep{1998MNRAS.301..640W} and with
the VLBA at 12.2~GHz (\cite{2000A&A...362.1093M},
\yearcite{2001A&A...369..278M}).
Our observation at 6.7~GHz provides the first VLBI image
for this source, in which fourteen spots were detected.
Twelve of which were clustered within 10~mas, the other two spots
were isolated \timeform{0".76} east from the main cluster.
These two clusters detected in our map correspond to 'M' and 'G'
defined by \citet{1998MNRAS.301..640W}, respectively.
There are several spectral features between 95 and 105~km~s$^{-1}$
in contrast to 12.2~GHz spectrum,
most of weak features could not be detected by our VLBI observation.

\subsection{IRAS~18556$ + $0136}\label{section:18556}
IRAS~18556$ + $0136 (figure~\ref{fig:fig8}) is at a distance of
2~kpc \citep{1982MNRAS.201..121B} and associated with
a CO outflows \citep{1985A&A...146..375D}.
Our VLBI observation is the first one for this source,
and six maser spots were detected.
This source consists of two clusters which are separated
larger than 5000~AU.
Although there are several spectral peaks, most of weak features
could not be detected.

\subsection{W~48}\label{section:W48}
W~48 (figure~\ref{fig:fig9}) is a well-known H\emissiontype{II} region
at a distance of 3.4~kpc \citep{1990A&A...233..553V}.
This source has a bright UC H\emissiontype{II} region that could be
an on-going massive star formation \citep{1989ApJ...340..265W}.
W~48 has been observed with EVN at 6.7~GHz \citep{2000A&A...362.1093M}
and with the VLBA at 12.2~GHz \citep{2000A&A...362.1093M}.
The maser of this source is strong and the spectrum is widespread,
24 spots forming a ring like structure were detected
as observed by \citet{2000A&A...362.1093M}.
A spot of 43.86~km~s$^{-1}$ at 34~mas north and 55~mas west from
the reference spot was newly detected one by this observation.

\subsection{OH~43.8$-$0.1}\label{section:OH43.8}
OH~43.8$-$0.1 (figure~\ref{fig:fig10}) is a star-forming region
at a distance of 2.8~$ \pm  $~0.5~kpc \citep{2005PASJ...57..595H}.
This star-forming region coincides with IRAS~19095$+$0930
which is an UC H\emissiontype{II} region \citep{1994ApJS...91..659K}.
This is the first VLBI observation at 6.7~GHz for this source
and we detected nine maser spots.
A spot of 43.00~km~s$^{-1}$ was located at \timeform{0".4} east
and \timeform{0".3} south from the reference spot.

\subsection{ON~1}\label{section:ON1}
Onsala~1 (ON~1, figure~\ref{fig:fig11}) is an UC H\emissiontype{II}
region located in the densest part of the Onsala molecular cloud
\citep{1983ApJ...266..580I}.
This source is known to be associated with a massive star-forming
region and IRAS~20081$+$3122.
A kinematic distance of 1.8~kpc is used by \citet{1998AJ....116.1897M}
and \citet{2004A&A...426..195K}.
Our observation is the first VLBI at 6.7~GHz
and seven maser spots were detected.
The redshifted cluster and blueshifted cluster are separated
about 1700~AU from each other.
It is surprising that the redshifted cluster corresponds to the narrow
($\sim$0.5~km~s$^{-1}$) spectral maximum with a flux density of 107~Jy,
the correlated flux density is only 4.6~Jy. The flux ratio of
correlated to total spectrum is 4.3~{\%}.
On the other hand, the blueshifted cluster correspond to
relatively weak ($\sim$20~Jy) and wide spectral peak,
while the flux ratio is about 50~{\%}.

\subsection{Cep~A}\label{section:CepA}
Cepheus A (Cep~A, figure~\ref{fig:fig12}) is a CO condensation
at a distance of 730~pc \citep{1957ApJ...126..121J}.
One of massive star forming regions in Cep~A is
an UC H\emissiontype{II} region, CepA-HW 2
\citep{1984ApJ...276..204H}.
Cep~A has been observed at 12.2~GHz with the VLBA
(\cite{2000A&A...362.1093M}, \yearcite{2001A&A...369..278M}),
while our observation is the first VLBI at 6.7~GHz.
We found 30 maser spots, 20 of which had $V_\mathrm{lsr}$ ranging
from $-$1.15 to $-$3.26~km~s$^{-1}$ and the other ten spots
had $V_\mathrm{lsr}$ ranging from $-$3.61 to $-$4.67~km~s$^{-1}$.
It is notable that spots in the redshifted cluster aligned
in the linear structure.
The blueshifted cluster is located at \timeform{1".34} east
from the redshifted cluster.

\subsection{NGC~7538}\label{section:NGC7538}
The NGC~7538 (figure~\ref{fig:fig13}) is a star forming region
at a distance of 2.8~kpc (\cite{1982ApJS...49..183B};
\cite{1984ApJ...279..650C}) including at least 11 high luminosity
infrared sources (NGC~7538~IRS~1$-$11),
which are probably young massive stars \citep{1990ApJ...355..562K}.
The central star has been thought as O6 (\cite{1976ApJ...206..728W};
\cite{1984ApJ...279..650C}), so the luminosity of the central source
is $ 8.3{\times}10^4~\LO $ and the mass is $ \simeq 30~\MO $. 
It is known that this object is associated with
an UC H\emissiontype{II} region that is observed with
the Very Large Array (VLA, \cite{1984ApJ...282L..27C};
\cite{1995ApJ...438..776G}).
The methanol maser of NGC~7538 has been observed with
the MERLIN \citep{2006A&A...448L..57P} and EVN at 6.7~GHz
(\cite{1998A&A...336L...5M}; \cite{2000A&A...362.1093M},
\yearcite{2001A&A...369..278M}; \cite{2004ApJ...603L.113P},
\yearcite{2006A&A...448L..57P}), and with the VLBA at 12.2~GHz
(\cite{1998A&A...336L...5M}; \cite{2000A&A...362.1093M},
\yearcite{2001A&A...369..278M}).
The distribution of maser spots in our map corresponds to IRS~1
region \citep{1998A&A...336L...5M}.
The map is in good agreement with that of \citet{2000A&A...362.1093M}.
The redshifted cluster spots with $V_\mathrm{lsr}$ ranging
from $-$56.45 to $-$55.75~km~s$^{-1}$ aligned in the linear structure
as well as previous studies (\cite{1998A&A...336L...5M};
\cite{2004ApJ...603L.113P}).
The spectral features of $-$53.07~km~s$^{-1}$ and $-$48.99~km~s$^{-1}$
corresponding to infrared sources IRS~11 and
IRS~9 \citep{2006A&A...448L..57P}, respectively, were not detected.


\section{Discussions}\label{section:discussion}
We have presented maps of twelve sources of methanol maser emission,
and the seven of them are the first VLBI results at 6.7~GHz.
The spatial distributions of maser spots show various morphology,
such as linear (Cep~A, NGC~7538), ring like structure (W~48),
largely separated clusters (W3(OH), W~33A, G~24.78$+$0.08,
G~29.95$-$0.02, IRAS~18556$+$0136, OH~43.8$-$0.1, ON~1, Cep~A).

We discuss the properties of methanol maser at 6.7~GHz as a possible
probe for astrometry comparing with the methanol maser at 12.2~GHz
and the water maser at 22~GHz.
We have detected twelve out of thirteen methanol masers at 6.7~GHz
with the longest baseline of 50~M$\lambda $ of our array.
The integrated flux density of the correlated spectra account for
$\sim$50~{\%} of the total flux,
and some of which account for more than 90~{\%}.
This high detection rate, flux recovery, and the small fringe spacing
of 4~mas suggest that most of the methanol maser emission
have compact structure.
This result is consistent with a previous study for W3(OH)
by \citet{1992ApJ...401L..39M}.
We also showed that the correlated flux density at 50~M$\lambda$
is typically 20~{\%} of the total flux density.
The size of maser spot inferred from the flux ratio
varies from 2 to 30~AU (at a distance from 0.73 to 9~kpc
of the sources).
This is consistent with the core size of 2 to 20~AU obtained
by \citet{2002A&A...383..614M}.
The velocity range of typically 10~km~s$^{-1}$
\citep{1995MNRAS.272...96C} is narrow compared to that of water masers.
Given this compactness and narrow velocity range as well as
the known properties of long-life and small internal-motion, 
this methanol maser line is suitable for astrometry with VLBI.

From the viewpoint of observation, atomospheric fluctuation is
a significant problem for observations at higher frequency.
Strong absorption by water vapor in the atomosphere makes
observations relatively difficult at 22~GHz in summer.
This would be a potential problem in measurement of annual parallax.
This is not the case for 6.7~GHz observation.

The astrometric VLBI observation uses continuum reference sources
which are usually distant quasars.
Such continuum sources show power-law, decreasing spectra,
and the flux density is larger at 6.7~GHz than
that of 12.2~GHz or 22~GHz.
This property makes the astrometric observations easier
in terms of detectability of reference sources.

The 6.7~GHz line might have some disadvantages due to
its lower frequency in comparison with the 12.2~GHz line:
The interstellar scintillation broadens the size of maser
and reference sources.
The interstellar broadening is stronger at lower frequency,
and might affect to measure the precise position of the sources.
Also the ionospheric density fluctuation changes path-length,
consequently affects phase measurement at lower frequency.
These effects depend on frequency as $\nu^{-2}$,
i.e., affect 3.3 times larger for 6.7~GHz than for 12.2~GHz.
Although \citet{2002A&A...383..614M} discussed that the
interstellar broadening is not significant at a scale larger
that 1~mas, we have to take account of these effects for astrometry.

We have made a phase-referencing VLBI observation with the JVN
and achieved that the positional accuracy of $\sim$50~$\mu $as
at 8.4~GHz and the image dynamic range of $\sim$50 on a target
\citep{2006PASJ...58..777D}.
The Bigradient Phase Referencing (BPR) was used for this observation,
and a reference calibrator was separated by a \timeform{2D.1}
from the target source.
The following equation (\ref{equ:equ1}) can be used to derive
an accuracy $\Delta \pi$ of annual parallax,
\begin{eqnarray}
 \Delta \pi = \frac{\theta_{\mathrm{beam}}}{D
              \cdot \sqrt{N_{\mathrm{spot}} \cdot N_{\mathrm{obs}}}}
\label{equ:equ1}
\end{eqnarray}
where $\theta_{\mathrm{beam}}$ is the minimum fringe spacing,
$D$ is Dynamic range of the image, $N_{\mathrm{spot}}$ is the number
of spots used in measuring annual parallax,
and $N_{\mathrm{obs}}$ is the number of observations.
If we make a series of observations for nine methanol maser spots
for five epochs as in the cases of the observation
by \citet{2006Sci...311...54X},
it is expected that an accuracy of $\sim$12~$\mu $as
in annual parallax would be achieved. 

The flux density at 6.7~GHz is typically $\sim $10 times larger than
that of 12.2~GHz \citep{1995MNRAS.274.1126C},
and the number of observable sources of 6.7~GHz is
much larger than that of 12.2~GHz.
This practical reason let us choose 6.7~GHz line than 12.2~GHz one.

We have started to improve the JVN at 6.7~GHz in terms of sensitivity
and the number of stations to detect weak masers and reference sources.
Usuda 64~m is one of the newly participating telescopes.
Assuming an aperture efficiency of 50~{\%} and $T_{\mathrm{sys}}$
of 50~K for all telescopes,
it is expected that sensitivity for fringe detection (7~$\sigma $)
would be less than 5~Jy.
We showed that the correlated flux density at 50~M$\lambda $ is
typically 20~{\%} of the total flux density.
Hence, sources with total flux density of larger than 25~Jy would be
potential targets for astrometric observation.
The number of such sources found in the catalog
by \citet{2005A&A...432..737P} is larger than 150.

\bigskip


The authors wish to thank the JVN team
for observing assistance and support.
The JVN project is led by the National Astronomical Observatory
of Japan~(NAOJ) that is a branch of the National Institutes of
Natural Sciences~(NINS), Hokkaido University, Gifu University,
Yamaguchi University, and Kagoshima University,
in cooperation with Geographical Survey Institute~(GSI),
the Japan Aerospace Exploration Agency~(JAXA), and the National
Institute of Information and Communications Technology~(NICT).


\end{document}